\documentclass[aps,prb,article,draftgroupedaddress,showpacs]{revtex4}
\usepackage{epsfig}  
\usepackage{graphicx}
\usepackage{amssymb}                               
\begin{document}

\title{Dynamic sound attenuation at hypersonic frequencies in silica glass}
\author{C. Levelut}  
\email{claire.Levelut@lcvn.univ-montp2.fr}
\author{R. Le Parc  and J. Pelous} 
\affiliation {Laboratoire des Collo\"{\i}des, Verres et Nanomat\'eriaux, CNRS/UMR5587,
Universit\'{e} Montpellier II, cc 69, 34095 Montpellier cedex, France}

\date{\today}

\begin{abstract}
In order to clarify the origin of  the dominant processes responsible for the
acoustic attenuation of phonons, which is a much debatted topic, we present Brillouin
scattering experiments in various silica glasses of different OH impurities
content. A large temperature range, from 5 to 1500 K is investigated, up  to 
the glass transition temperature. Comparison of the hypersonic wave attenuation in
various samples allows to identify two different processes. The first one induces a  low
temperature peak related to relaxational processes; it 
 is strongly sensitive to the extrinsic defects. The second, dominant in the high
temperature range, is weakly dependent on the impurities  and can be ascribed
 to anharmonic interactions.

\end{abstract}


\maketitle

The origin of sound attenuation in glasses has been a highly 
debated topic in recent years. Several models have been 
proposed  but  no consensus about the interpretation of 
sound attenuation  has been reached.\cite{Benassi1996,Foret1997,Benassi2005}

Several mechanisms responsible for sound attenuation have been 
identified for over the year:
	
\begin{enumerate}
\item interactions of acoustic waves with defects characteristic
 of the disorder. Those defects are often described as tunneling 
 systems  responsible for many anomalous properties of glasses in 
 the low temperature regime.\cite{HunklingerArnold1976}

\item thermally activated relaxational processes, as clearly demonstrated
 by ultrasonic measurements.\cite{Anderson1955}  The microscopic 
 structural origin of the thermally activated processes is often
  related to tunneling defects or soft modes responsible for the 
  interaction with acoustic waves.\cite{Buchenau1992}

\item Rayleigh-like scattering by static inhomogeneities, 
independent of temperature. This has been 
observed for example in porous glasses (xerogels, 
aerogels).\cite{Courtens1987,Caponi2004} This mechanism yield an attenuation 
varying like  $q^4$, where $q$  is the momentum transfer of the interaction in scattering
experiment. 
 Rayleigh-like scattering has been also 
put forward  to explain the change of regime for the attenuation 
in the THz range.\cite{Benassi2005,Rat1999} An 
alternative explanation of the frequency power dependence in 
$\omega^4$ (where $\omega=2\pi/\nu$ is the pulsation of the excitation) 
in terms of fractons\cite{Alexander1986} has 
been also proposed.

\item anharmonicity responsible for the  interactions of acoustic 
waves with thermal phonons, as in crystals.\cite{Vacher1981,Fabian1999}
\end{enumerate}

Each mechanism dominates in different temperature and frequency ranges. 
It is expected that the mechanisms 1 and 2 are dominant at low 
temperatures, 3 in very inhomogeneous media or at  wavelengths  comparable
 to a few atomic length  and the mechanism 4 at high temperature and 
 high frequencies.
Moreover, the influence of several parameters have been studied in 
order to distinguish between universal behavior of the sound 
attenuation in glasses and more specific properties, related 
for example to composition effects,  or to the method of preparation 
(impurity concentration or thermal history).\cite{Krause1964} The  influence of
 pressure, \cite{Tielburger1992} 
irradiation by neutrons\cite{Bonnet,Laermans} or permanent 
densification \cite{Rat1999} has also been
investigated.

In this paper we present results about Brillouin scattering 
measurements of sound attenuation at hypersonic frequencies 
in silica. This work provide new data useful for shedding  new 
light on the origin of processes responsible for the sound 
wave attenuation. By comparing results for silica with different
 impurity content, we will  demonstrate that in the hypersonic regime: 

\begin{itemize}

\item the thermally activated relaxational processes  which dominate
 at low temperature are clearly related  to specific, clearly 
 identified extrinsic structural defects; 
\item at high temperature, anharmonic processes are less dependent
 on these local changes of composition.
\end{itemize}

Two samples of vitreous silica of different origins were 
investigated. One of them is a commercial fused quartz ``puropsil'') 
provided by Quartz et Silice, Nemours, France. This sample contains
 a few ppm (less than 20) of OH impurities. Another   sample 
 was prepared by densification of a silica aerogel. The sample 
 of initial density \mbox{0.3 g cm$^{-3}$}, was heat-treated at  about 
 \mbox{1100 $^{\circ}$C}  for several hours till its density reaches 
 the  same value, \mbox{2.20 g cm$^{-3}$}, as 
   amorphous silica. This sample has both a 
 different "fictive temperature"(lower than \mbox{1100 $^{\circ}$C} compared 
 to the commercial samples   (between 1200 and \mbox{1300  $^{\circ}$C}) and a 
 different OH content  (around 3000 ppm,  determined with
 infrared spectroscopy). In the following, it will be referred
  to as the sol-gel sample.
We used Brillouin scattering measurements performed with a
 high resolution Fabry-P\'erot (FP) spectrometer.\cite{VacherSussner1980} 
 The incident  light 
 was the  514.5 nm  line of a single-mode 
 argon-ion laser (Spectraphysics 2020) with 
 an incident  power on the  sample of about 1W.  
 This set-up comprises a double-pass plane FP, 
 whose free spectral range is equal to 75 GHz  and 
 finesse equal to 40, used as a filter (monochromator),
  followed by a spherical FP interferometer with a free
   spectral range of \mbox{1.48 GHz} and a finesse of 50 as 
the   resolving unit. The measurements of the longitudinal 
   sound wave velocity and attenuation are performed in 
   backscattering geometry.  The accuracy of the 
   experimental data for the sound velocity, deduced from the Brillouin shift,  and the 
  sound attenuation, related to Brillouin linewidth, are respectively of about  0.1\% and  5\%.
    Low temperature measurements were performed using a 
    He cryostat. High temperature measurements were carried out 
    using a Hermann-Moritz oven with optical windows, which 
    allows measurements  up to \mbox{1500 $^\circ$C }.The data are analyzed 
    using a non-linear
    fitting procedure. The Brillouin profiles were adjusted with a Lorentzian function of 
    half width at half maximum $\Gamma$ convoluted with  an apparatus function measured using the elastic line
  of an experimental spectrum.
The  internal friction  parameter $Q^{-1}$ is  deduced 
from the  the full width at half maximum   $2\Gamma$ and 
 the position $\delta \nu$ of the longitudinal Brillouin line, using:
$Q^{-1} =  2\Gamma/\delta \nu$.

\begin{figure}
\centerline{\epsfxsize=260pt{\epsffile{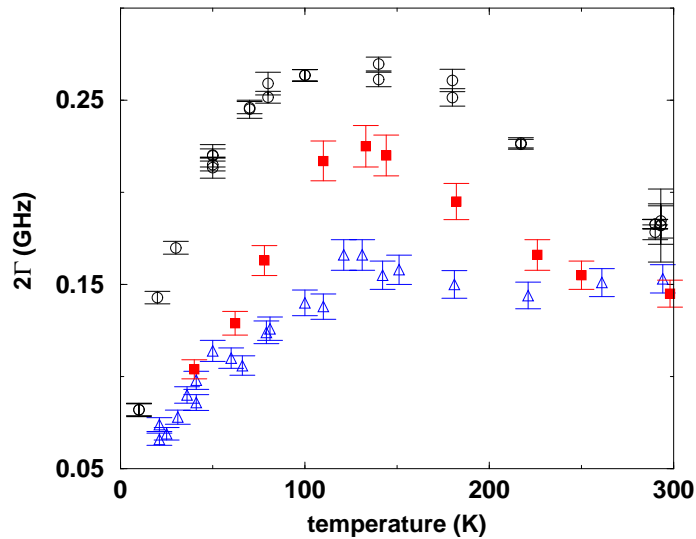}}}
\caption{\label{fig:fig2} Full width at half maximum of the Brillouin peak
  in a three silica samples plotted as  functions of temperature from 5 to
300 K for the sol-gel sample ($\circ$) ([OH]=3000 pmm), for suprasil\cite{Pelous1978} 
 ($\square$)
  ([OH]=1200 ppm) and for puropsil\cite{Vacher1976}   ($\vartriangle$) ([OH]=20 ppm). }
\end{figure}
 
\begin{figure}
\centerline{\epsfxsize=250pt{\epsffile{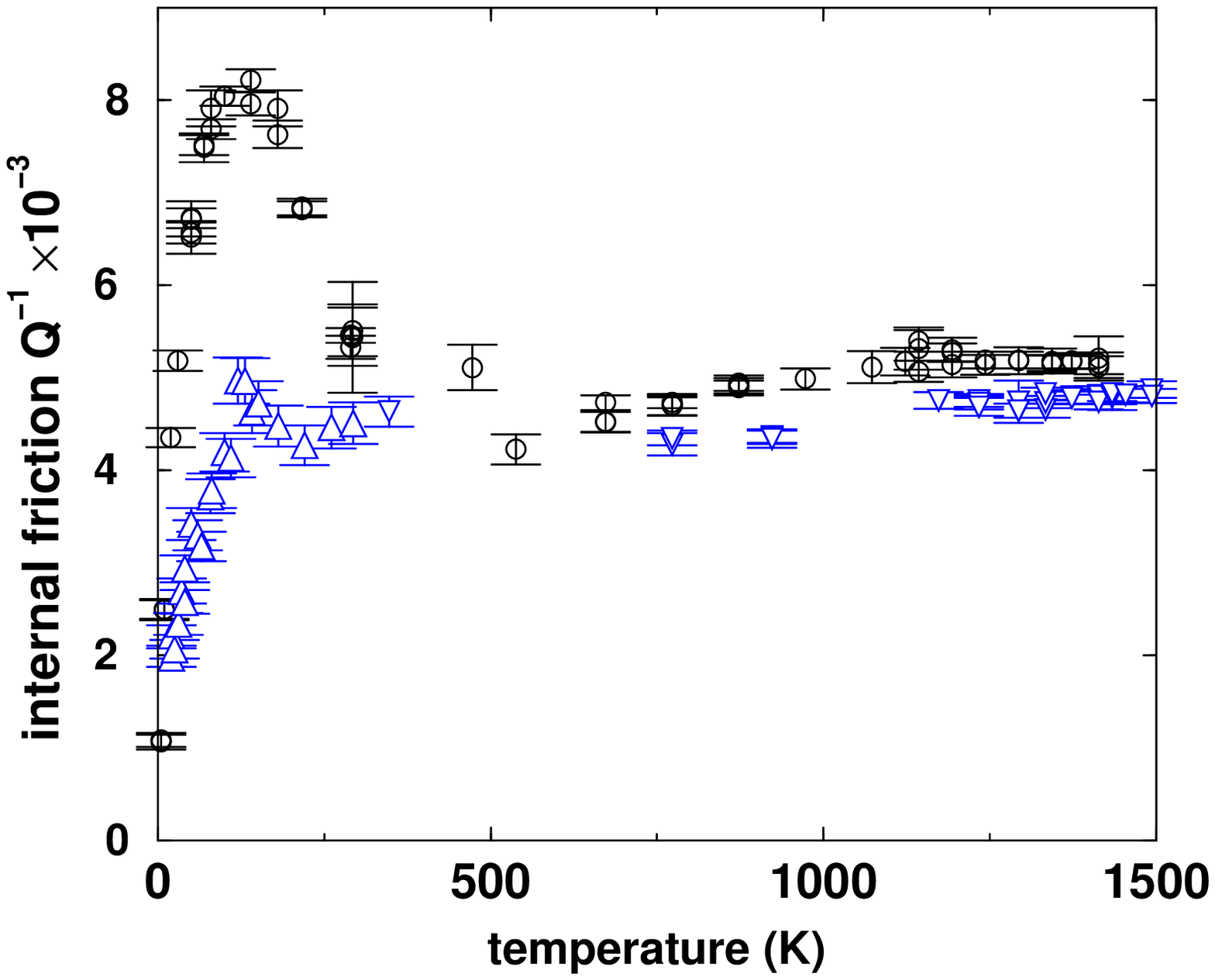}}}

\centerline{\epsfxsize=250pt{\epsffile{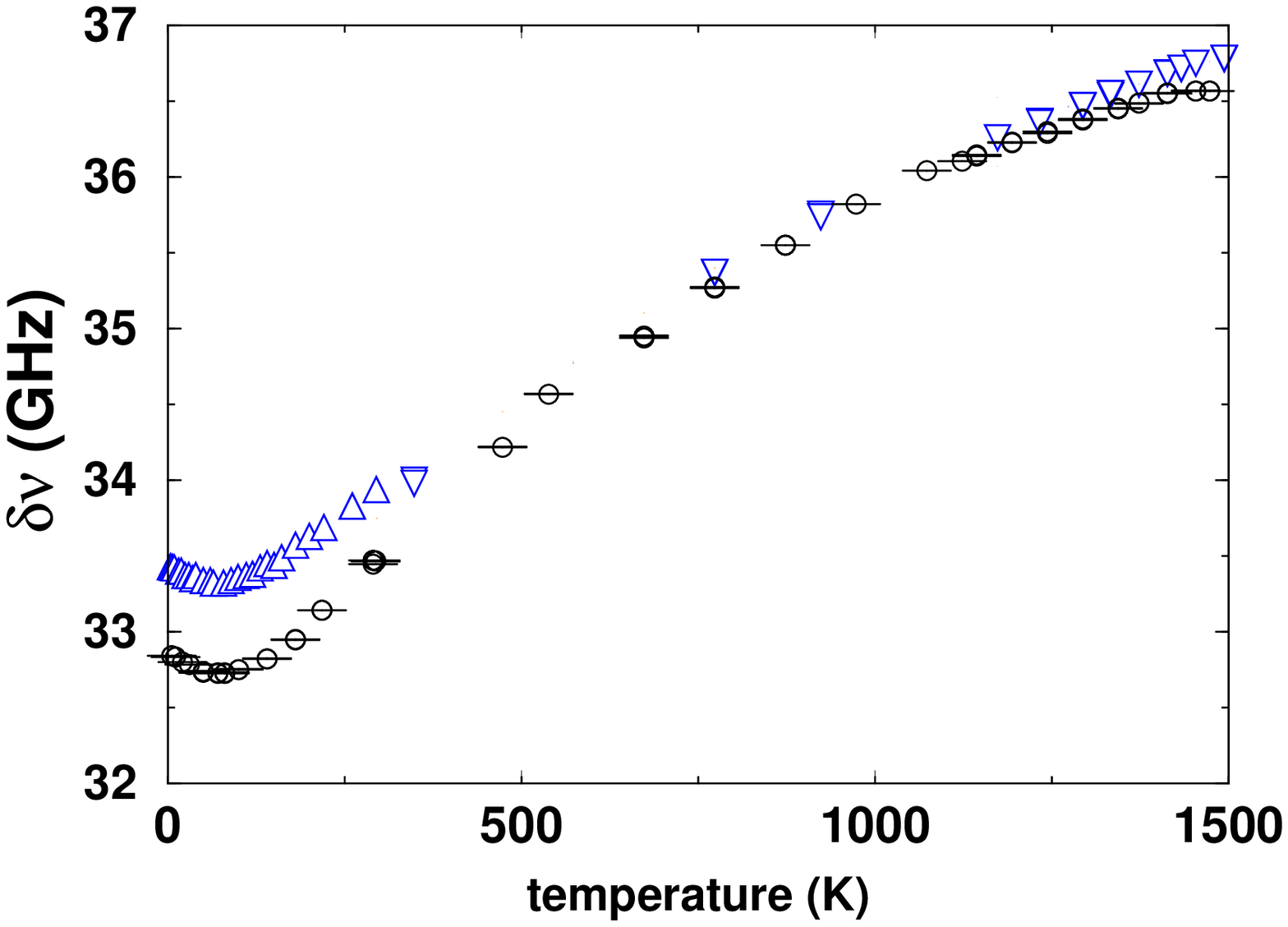}}}

\caption{
a) Internal friction in a two silica samples (sol gel
($\circ$) and puropsil ($\vartriangle$ ($\vartriangle$ for low temperature data \cite{Vacher1976},
$\triangledown$ for high temperature new data), plotted as a function of temperature, from 5 to
1500K. The results for the entire temperature range for the sol-gel silica and above room
temperature for the puropsil are new results. The low temperature results for puropsil were
 published previously.\cite{Vacher1976}. b) Brillouin frequency shift
  in sol-gel silica samples ($\circ$), plotted as a function of temperature, from 5 to
1500 K, compared to data for a low OH content silica glass, puropsil ($\vartriangle$,$\triangledown$). The low temperature 
measurements are taken from
Ref. ~\onlinecite{Vacher1976}, the high temperature ones are new results.  The error bars 
account for the  error in the fitting procedure,
 an aperture  of $\pm 1^{\circ}$  for the collected scattering, an error 
 in the determination of the position of the elastic lines and of the thickness of the
 Fabry-P\'erot. \label{fig:fig3}}

\end{figure}

 The
linewidth in the region of the minimum for  sound velocity, from 5 to 300K,  is plotted in  
Fig. \ref{fig:fig2}
 for the sol-gel silica, as well as for two 
 commercial silica samples previously investigated (suprasil and
puropsil)\cite{Vacher1976,Pelous1978}. 
The internal friction is  shown on  Fig.
 \ref{fig:fig3}a. 
  The results for $Q^{-1}$ in puropsil are  
  very similar to results in another commercial sample (suprasil
  W characterized by a very small amount of impurities), measured using a different wavelength
  (\mbox{$\lambda=$ 488 nm}).\cite{Tielburger1992} 
 The attenuation versus temperature curves exhibit a large peak at
   low temperatures, centered around \mbox{150 K}, as first observed at hypersonic 
   frequencies long ago\cite{Pine1969} and then a nearly constant 
   value up to the glass transition temperature range.  The Brillouin frequency shift
    for sol-gel silica  and puropsil are presented
 from 5 to \mbox{1550 K} in Fig.
 \ref{fig:fig3}b. The frequency shift is needed to calculate the dimensionless internal
 friction parameter, but also to determine sound velocity. The data of Fig. \ref{fig:fig3}b 
  are close to
 previous determination of the literature in several commercial silica
 samples, \cite{BucaroDardy1974,KrolLyons1986,Polian2002} and the high
  temperature part of
 those curves will  be also discussed,
 in comparison with previous determinations,  in another article (Ref. ~\onlinecite{LeParc2005}).

Two main features can be deduced from the figures.
 First, the attenuation peak around 150K is more
 prominent in  silica samples containing more OH impurities and having lower fictive temperature.
 Taking the  observed  high temperature plateau
as a reference level, the peak height for the linewidth 
 in sol-gel silica is about 0.12 GHz vs 0.07 GHz in
 suprasil, whereas it is very weak (around 0.015 GHz) in puropsil sample
  which contains very few OH
 impurities. A similar increase of the attenuation peak can also be 
 observed in Fig. 2 of Ref.
 ~\onlinecite{Caponi2004} where the authors report measurements 
 of the attenuation over a limited temperature
 range in a densified
 aerogel, of density 2.19 g cm$^{-3}$ and  whose OH content is unknown. 
 Second, within the 
 accuracy of the experiments the values of 
internal friction above room temperature are 
almost identical in the low and high OH content silica, in fact they are slightly higher 
for the sol-gel silica 
compared to commercial  samples. Variations of the sound velocity with the OH content
 can also be observed on Fig. \ref{fig:fig3}b, mainly in the range of the attenuation peak and
 of the glass transition.

It has been proved previously that the low temperature
 peak has the same origin as the peak observed at lower 
 temperature using lower frequencies, such as ultrasonic 
 measurements. Several authors\cite{Pine1969,Vacher1976, Tielburger1992} 
   have checked that the temperature variation of the 
   peak position is in agreement with  thermally 
   activated processes 
   having a distribution of activation
    energy and  a mean activation energy in the \mbox{300-600 K}
    range depending on the form of the distribution of 
    energy barriers.\cite{Bonnet1991}
Using ultrasonic investigation, many studies have 
been devoted to the influence of the preparation 
method \cite{Krause1968,Krause1971} and of the impurity content 
(for example OH impurities) on the amplitude of this 
peak. It can be pointed out that such a peak is 
observed in many glasses. As examples, the 
ultrasonic attenuation at \mbox{20 MHz} exhibit a
 peak around \mbox{170 K} in GeO$_2$ and around \mbox{30 K} in BeF$_2$ 
 \cite{Krause1968,Strakna1964}. In B$_2$O$_3$ the 
 amplitude and the temperature position of the peak have 
 been strongly related to OH content.\cite{Kurkjian1966} 
 The main difference between the  three silica samples whose 
 attenuations are presented in Fig. \ref{fig:fig2} is the OH content 
 (20 for puropsil, 1200  ppm for suprasil and  \mbox{3000 ppm} for sol-gel silica).
  Therefore, this result demonstrates, 
 for the first time at hypersonic frequencies, that the 
 impurities are involved in this feature of the acoustic 
 attenuation. This result does not mean necessarily that 
 OH are  really involved in the thermally activated 
 jumping. OH defects can induce local structural
  modifications \cite{Kakiuchida2002} and create 
  entities participating in jumping. Indirect structural effects
   can also be due to a decrease 	of fictive temperature  induced by
   increasing the OH content.  
  Nevertheless, 
  such a damping peak at hypersonic frequencies
   related directly to the presence of OH impurities 
   has also been observed in crystals such as KCl (Ref.~\onlinecite{Berret1983}).
    Observed also by infrared absorption, Raman scattering 
    and thermal conductivity measurements, this relaxational 
    process has been attributed to librational motions of 
    the center of mass of the hydroxyl molecule. This 
    observation in crystals suggest that a direct 
    implication of OH cannot  be excluded in glasses also.

   If the attenuation peak is related to thermally activated processes, an associated change on the temperature
    dependence of velocity should is also expected. Such variation is observed on Fig. \ref{fig:fig3}b). 
    It has been pointed out, using ultrasonic 
measurements, \cite{Krause1971} that the fictive temperature does not modify significantly 
 the temperature dependence of velocity in the region of the
attenuation peak (around 80K). The temperature variation of sound velocity in this range depend only on the OH content.
 Thus we could expect that the attenuation in this range depend more strongly on the OH content than on the fictive temperature.
Figure 1 of Ref. ~\onlinecite{Krause1971} shows that the influence of fictive temperature on the
heigt of the damping peak  at hypersonic frequency decreases with increasing OH content. 
We expect the same tendency at hypersonic frequency ant then the amplitude of the peak
in sol-gel silica should be in very large part  due to high OH content. Moreover, we
 have measured the fictive temperature influence at room temperature on the linewidth of a very low OH content silica glass, and we found that a variation of
 fictive temperature by 400 K induces only a decreases of about 10\% in the linewidth (from 0.144 GHz for $T_f=1373$K to 0.159 GHZ for a sample with
 $T_f=1773$K) or in the internal friction ($4.23 \times 10^{-3}$ to $4.7 \times 10^{-3}$).

In contrast, the presence of impurities seems 
to have  little effect on the friction coefficient
at temperatures higher than \mbox{300 K}. Above room temperature, 
the attenuation exhibits a plateau over a wide temperature 
range (up to more than \mbox{1500 K} for the OH free sample). In  
contrast to  the attenuation peak region, in the high 
temperature range, the hypersonic attenuation is 
comparable to that measured in crystals.\cite{Pelous1976}
 Moreover it has been pointed out that the attenuation in 
 permanently densified silica is also very close to that 
 of crystalline quartz. \cite{Rat2005} As the relaxational 
 process can not explain the amplitude or the frequency 
 dependence in $\omega^2$ of the attenuation in the high temperature 
 range,  \cite{Zhu1991} anharmonicity has been invoked
  previously as the dominant process involved in the
   hypersonic attenuation at high temperature in silica 
   glass.\cite{Vacher1981,Foret2005} Recently new 
   theoretical approaches have been proposed to describe 
   some aspects of anharmonicity in glasses\cite{Gurevich2003,Novikov1998} but up 
    to now the applicability of these models to 
    quantitative analysis of hypersonic attenuation 
    is lacking. Using the framework of the same
     formalism as for crystalline materials  \cite{Maris1971} the anharmonic damping 
     can be understood as a coupling of hypersonic 
     waves with the bath of thermal phonons. Indeed,
      simulations\cite{ Fabian1999} have shown that 
      anharmonicity process can induce an anomalously large 
     Gr\"uneisen parameter and sound attenuation due to 
   larger  anharmonicity  in glasses than in crystals. 
     Quantitative analysis in terms of anharmonicity has
      been recently successful\cite{Foret2005} in describing
       hypersonic attenuation in OH free silica glass up 
to 300K. To follow the same analysis in a OH rich 
sample, the influence of local impurities on the 
phonon vibration  spectra  should  be examined.  
It is expected and verified by experiments that 
local intrinsic disorder due to impurities does not
 contribute to the large variation of the vibrational
  spectra\cite{Geissberger1983}. So, at
  high temperature when the population of high 
  energy phonons is significant the mean thermal
   relaxation  of  phonons is little modified by
	  the impurities. Therefore, in the framework
	    of the interpretation
    of an attenuation due to the interaction of 
     hypersonic waves with the whole thermal phonon bath, it follows 
	     that the attenuation in this frequency and 
	   temperature range is only weakly affected by 
	 impurity content as shown in our experiments.

The main contribution to acoustic attenuation in silica glass, 
measured at hypersonic frequencies  by Brillouin scattering 
 for temperature lower than 300K,  is a  large relaxation 
 peak due to  an interaction with structural defects strongly 
 related to extrinsic impurities such as OH impurities. 
  Comparison of acoustic attenuation at various 
 frequencies are  often made in the literature. However, the  
  large variations of the damping amplitude which could occur 
   due to the 
 relaxational process related to impurities, require that the origin and 
 impurity content of
 the  samples under comparison should  to be taken into account.
The relaxational contribution, due to thermally 
activated processes, dominates only at low 
temperatures and does not account for the high 
temperature regime of the attenuation where anharmonicity
 is likely to be the dominant process.

\section*{Acknowledgment}
The authors
 express their acknowledgments to  T. Woignier for
 the preparation of a high optical 
quality sol-gel silica. We thank  R. Vacher and  R. Vialla who designed and built the 
 high resolution Brillouin spectrometer used in this work. We also thanks I. Campbell for
 critical reading of the manuscript.
\bibliography{art_ATTENUATION}
\bibliographystyle{apsrev}

\end{document}